\documentclass[12pt]{article}
\usepackage[T1]{fontenc}
\usepackage[utf8]{inputenc}
\usepackage[absolute,overlay]{textpos}
\usepackage{hyperref}
\usepackage[numbers,sort&compress]{natbib}
\usepackage{authblk}
\usepackage{blindtext}
\usepackage{graphicx}
\usepackage{amssymb}
\usepackage{titlesec}
\usepackage{placeins}
\usepackage{xcolor}
\usepackage[all]{nowidow}
\usepackage[title]{appendix}
\usepackage[top=1.5in, left=1.5in]{geometry}

\newcommand*\diff{\mathop{}\!\mathrm{d}}

\begin{document}

\title{Application of space-time spectral analysis for detection of seismic waves in gravitational-wave interferometer}

\author[1]{Robert Szymko}
\author[2]{Mateusz Denys\footnote{e-mail: mateusz.denys@gmail.com}}
\author[2,3]{Tomasz Bulik}
\author[3]{Bartosz Idźkowski}
\author[3]{Adam Kutynia}
\author[3]{Krzysztof Nikliborc}
\author[3]{Maciej Suchiński}

\affil[1]{\small Faculty of Physics, University of Warsaw, Pasteura 5, PL-02093 Warsaw, Poland}
\affil[2]{\small AstroCeNT --- Particle Astrophysics Science and Technology Centre International Research Agenda, Nicolaus Copernicus Astronomical Center, Polish Academy of Sciences, Rektorska 4, PL-00614 Warsaw, Poland}
\affil[3]{\small Astronomical Observatory, University of Warsaw, Ujazdowskie Ave. 4, PL-00478 Warsaw, Poland}

\maketitle

\begin{abstract}

Mixed space-time spectral analysis was applied for the detection of seismic waves passing through the west-end building of the Virgo interferometer. The method enables detection of every single passing wave, including its frequency, length, direction, and amplitude. A thorough analysis aimed to improving sensitivity of the Virgo detector was made for the data gathered by 38 seismic sensors, in the two-week measurement period, from 24 January to 6 February 2018, and for frequency range 5--20 Hz. Two dominant seismic-wave frequencies were found: 5.5 Hz and 17.1 Hz. The possible sources of these waves were identified, that is, the nearby industrial complex for the frequency 5.5 Hz and a small object 100 m away from the west-end buiding for 17.1 Hz. The obtained results are going to be used to provide better estimation of the newtonian noise near the Virgo interferometer. 

\end{abstract}



\section{Introduction}
\label{intro}

First events detected by gravitational-wave (GW) detectors in the past decade gave us hope to answer the most fundamental questions of astronomy and physics, concerning, e.g., evolution of galactics and black holes, gravitational colapse of supernovae, or limits of general relativity \cite{abbott2016, sesana2016, endlich2017, sasaki2018, miller2019, abbott2019}. However, noise reduction in these detectors is most complex task, due to very low amplitude of the incoming signals. Seismic characterization of the vicinity of the GW detectors is necessary for the noise reduction, especially for filtering \emph{newtonian noise} coming from fluctuating gravitational forces caused by density perturbations in the Earth \cite{saulson1984terrestrial, hughes1998seismic, couglin2016}

\subsection{Space-time spectral analysis}

Space-time spectral analysis (or the frequency-wavenumber method) is a straightforward extension of harmonic analysis to more than one dimension. Apart from a temporal dimension, additional spatial dimensions are introduced. It is most suitable if the spatial dimensions are cyclically continuous, or at least have fixed boundaries. In such cases we can look for modes of variability in which spatial scales have particular temporal scales. If the behavior is harmonic (wavelike), then we expect space-time spectral analysis to isolate any such modes that are present in the data. For instance, if we make space-time spectral analysis of a stringed instrument, we expect to find a definite relationship between the length scales and the time scales of the oscillations \cite{hartmann2008objective}. 

The method presented in this article is based on the paper \cite{lacoss1969estimation} (and refs.\ therein), where seismic noise structure has been estimated using arrays of detectors (see also \cite{gazdag1984migration}). This method is one of the most popular methods of seismic-wave detection to the present day \cite{sato2012seismic, foti2014surface, foti2011application}. It is straightforward and provides a lot of useful information about the seismic activity at the same time. Some more complex alternatives to this method are, e.g., High-Resolution method of Capon \cite{capon1969}, the broad-band frequency-wavenumber spectral method of Nawab \emph{et al.} \citep{nawab1985}, or the Multiple Signal Classification scheme \cite{schmidt1986} (see also \cite{rost2002, litehiser2018}).

Space-time spectral analysis has been used before to obtaining the Rayleigh wave phase velocity dispersion curves \cite{mainsant2012ambient}. It is also a widely explored method in the atmospheric sciences \cite{kao1968governing, hayashi1982space, salby1982sampling, wu1995least, hendon2008some}.\footnote{That is, a version with one spatial dimension plus temporal dimension.} In \cite{hayashi1977space, hayashi1979generalized} space-time spectral analysis has been used for the description of zonal component of winds over the equator. Moreover, \cite{de2005spectral} applied space-time spectral analysis to modeling rainfall fields. As far as only spatial Fourier transform is concerned, Kim and Barros \cite{kim2002space} used its two-dimensional version to analyse structure of soil moisture fields in different time moments, considered separately. 

Investigations of the seismic characteristics of the Virgo site have been made previously in \cite{acernese2004properties, saccorotti2011seismic}. For some recent works about seismic-noise measurements for gravitational-wave detectors see, e.g., \cite{somlai2019seismic, van2019long, mukund2019ground, tringali2019seismic}. 

This paper is structured as follows. The system of seismometers that we used for data aquisition is presented in Sec.\ \ref{system}. Then, we provide information about data analysis process (Sec.\ \ref{method}). The results of this analysis are briefly described in Sec.\ \ref{results}. Finally, some concluding remarks are given in Sec.\ \ref{summary}.

\section{Measuring system}
\label{system}

\begin{figure}
\centering
\includegraphics[width=0.9\textwidth]{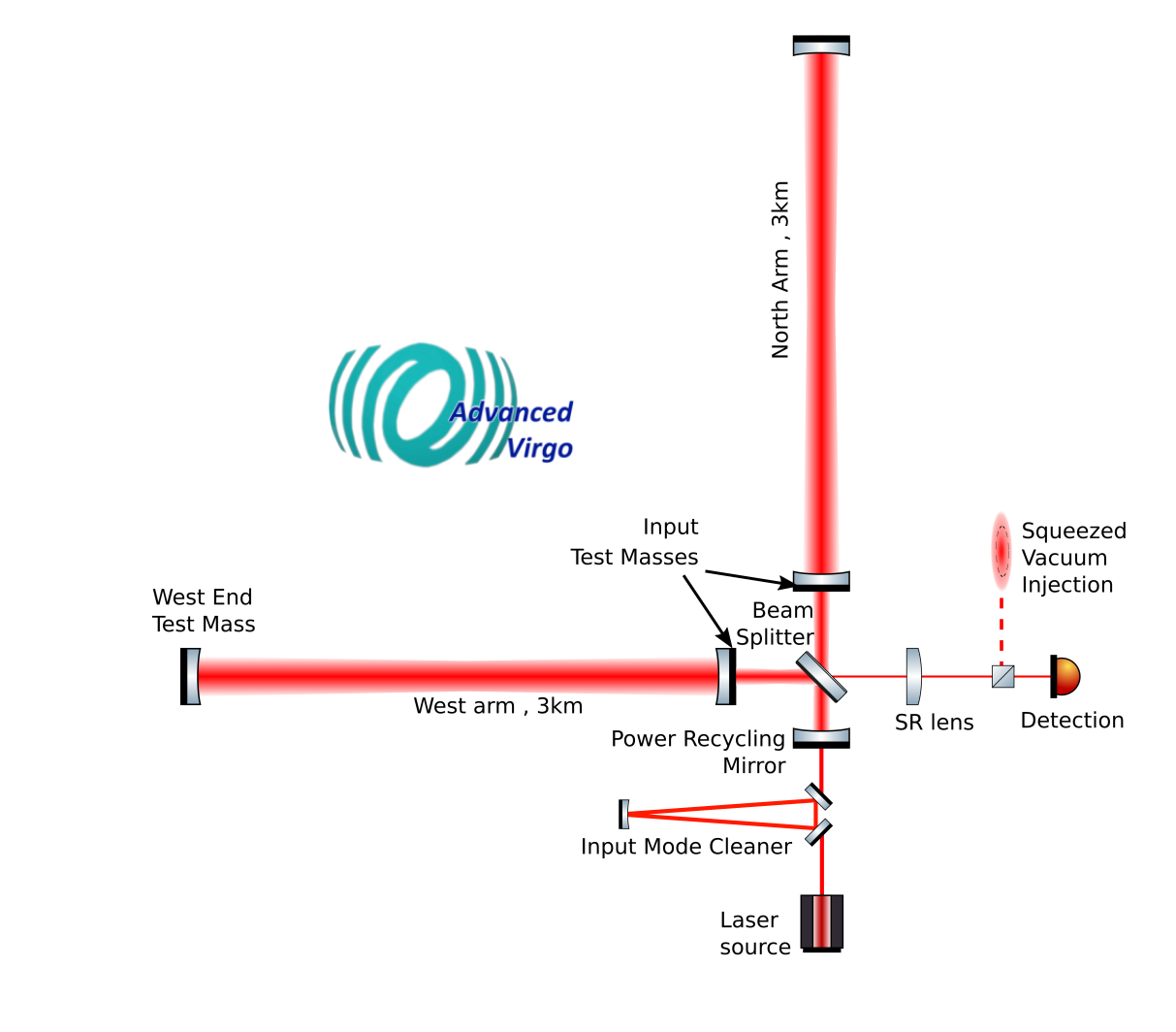}
\caption{Advanced Virgo layout.\protect\footnotemark}
\label{fig:Virgo_schem}
\end{figure}

\begin{figure}
\begin{center}
\includegraphics[width = 0.8\textwidth]{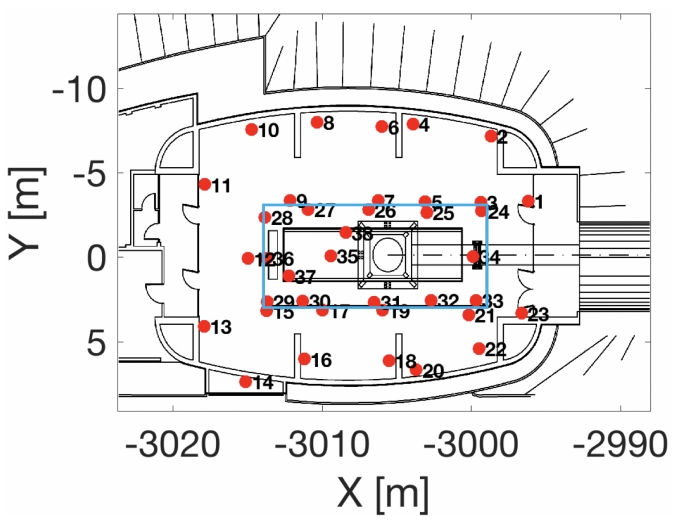}
\caption{Location of the 38 seismometers in the west-end building (taken from \cite{tringali2019seismic} by courtesy of Tomasz Bulik).}
\label{fig:latt}
\end{center}
\end{figure}

During the study a system of seismic detectors in the west-end building (WEB) of the Virgo interferometer was examined. Location of the WEB as a part of the Virgo detector is shown on Fig.\ \ref{fig:Virgo_schem}.
\footnotetext{http://public.virgo-gw.eu/advanced-virgo/}
The system consists of 38 seismometers, located inside the building as shown on Fig.\ \ref{fig:latt}.
The seismometers no.\ 1--23 were placed on the floor of the building, those no.\ 24--38 --- on the central platform, while those no. 37 and 38 --- in the basement. The floor is supported by 30 m long poles, while the central platform and the basement are supported by a set of longer poles reaching a more stable gravel layer, at 52 m depth. There is a vacuum chamber on the central platform where the Virgo test mass is located, together with a suspension and a vibration isolation system. The construction also provides suppression of seismic noise above 15 Hz on the central platform. Therefore, for frequency values above ca.\ 15 Hz, there is a significant difference in the amplitude spectral densities (ASDs) between the platform and the floor \cite{tringali2019seismic}. 

The sensors were placed on the floor fixing their heavy mount plate with double-sided adhesive tape for a good connection to ground, which is essential for low-temperature measurements. Each sensor contains also a preamplifier and an analog-to-digital converter to avoid coupling with electromagnetic noise when the signals are transmitted through the cables surrounded by other electrical devices. For more details see \cite{tringali2019seismic}. 

The exact coordinates of the seismometers are provided in Appendix \ref{appendix}. The average distance between the seismometers was approximately 9.5 m. Point $(0,0)$ of the coordinate system shown on Fig.\ \ref{fig:latt} is a beam spliter in the central building. The axes coincide with the arms of the detector, therefore, the axes form an angle ca.\ $30^{\circ}$ with cardinal directions. 

\begin{figure}
\centering
\includegraphics[width = 0.8\textheight, angle=270]{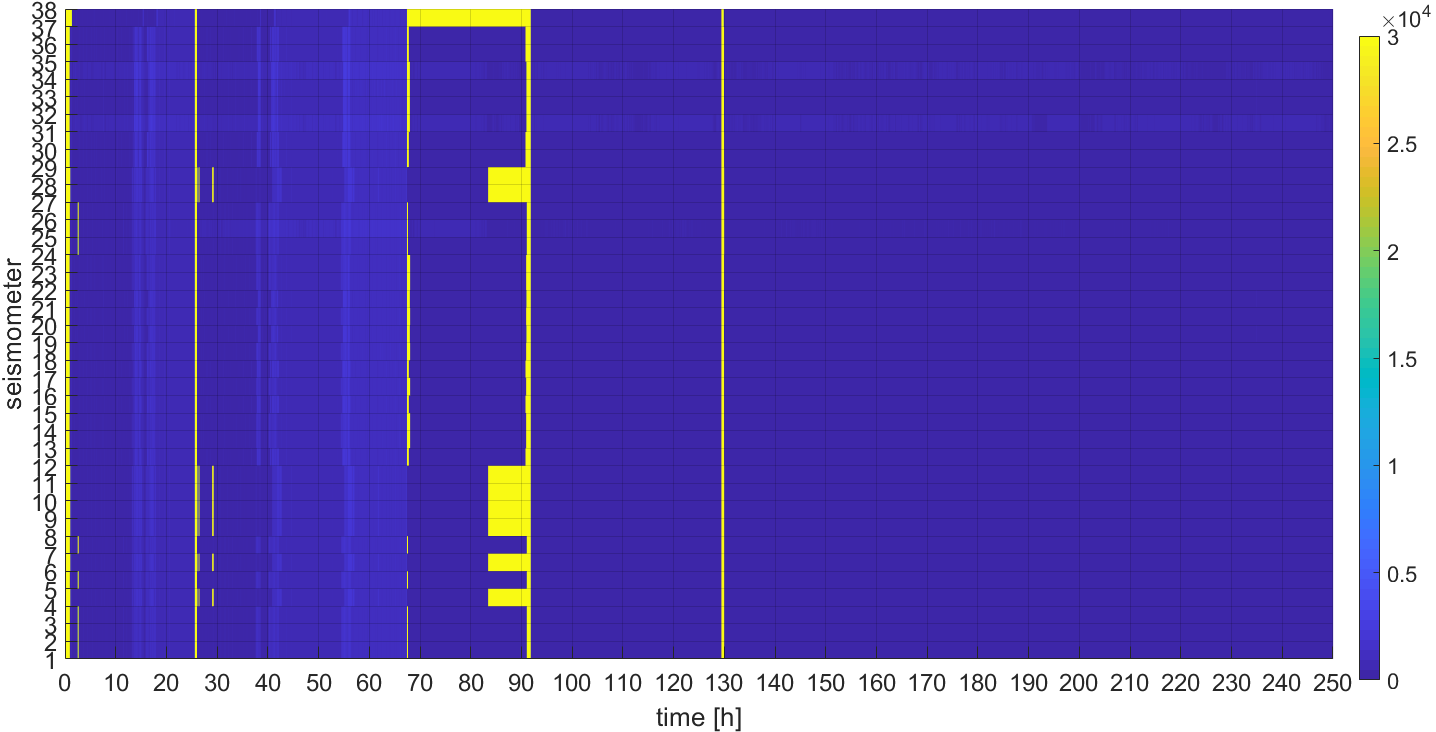}
\caption[cz]{Quality of the data. Yellow color stands for measuring 'zero' values by the sensor for the whole minute (30 000 measurements). Whiter shades of blue corresponds to smaller lacks of data. For hours: 1, 26, 68, 92, 130 there are empty periods for all sensors. Between 83 and 92 hour 9 seismometers (no.\ 4, 6, 8, 9, 10, 11, 27, 28, 38) had large empty period --- these seismometers are localized on the north side of the mirror (cf.\ Fig.\ \ref{fig:latt}).}
\label{fig:quality}
\end{figure}

The data that were used for the calculations were collected during a two-week measurement period, from 24 January to 6 February 2018. The data for each detector are values of voltage changes due to its vertical motion, collected with the sampling frequency 500 Hz, continuously. Thus, the system measures surface Rayleigh waves. Not all (256) 1-hour data files generated in the measurement period are complete. Actually, the real measurement time is 250 h ($\sim 10.5$ days). During the experiment a series of each sensor's 10 'zero' measurements one after another was assumed being empty-data period for this sensor. Collectively, this is $3.4\%$ of measurement time for all sensors. On Fig.\ \ref{fig:quality} empty periods for all sensors are presented. The breaks in data collection for all sensors are not showed, therefore, hours on the horizontal axis do not correspond to real time. 

Although the sensors measure ground motions, actually the system was created in order to estimate newtonian noise in the Virgo measurement signal, i.e., a noise coming from fluctuating gravitational forces caused by density perturbations in the Earth, which can be estimated using seismic-noise signal \cite{saulson1984terrestrial, hughes1998seismic}. Newtonian noise has to be subtracted from the signal in any ground-based gravitational-wave detector (cf.\ \cite{van2019long, harms2015newtonian}). The goal of this analysis is to improve sensitivity of the Virgo for low frequencies.

\section{Methodology}
\label{method}

For the above mentioned data, the following formula for the mixed space-time Fourier transform was applied,
\begin{equation}
\label{STFT}
\textrm{FT}(k_x, k_y, \omega) = \int_0^t \int_{-\pi}^{\pi} \int_{-\pi}^{\pi} e^{-2 \pi \mathit{i} (x k_x + y k_y + t \omega)} \psi(x, y, t) \diff x \diff y \diff t, 
\end{equation}
where $\psi(x, y, t)$ is a measurement value of the seismometer placed in $(x, y)$\footnote{Because the system of seismometers measure surface waves, $z$ axis is not included in the above formula.} in time moment $t$, $\omega$ is a frequency of the detected wave, and $k_x, k_y$ are the respective components of its two-dimensional wave vector $\textbf{k}$. Actually, a discretized version of Eq.\ (\ref{STFT}) was used for $\Delta t = 0.002$ s. The value of $\textrm{FT}(k_x, k_y, \omega)$ provides information about the amplitude of the particular wave. Notably, both spatial dimensions should not exceed $(-\pi, \pi)$ range, so they should be normalized to that range (see below). 

In general, the space-time Fourier transform consists of wave vectors for each wave detected in the signal ($k_x, k_y$ axes) together with the frequency of that wave (on $\omega$ axis). Thanks to that we can obtain length, direction, frequency and amplitude\footnote{By calculating an inverse Fourier transform of each ASD peak in the final result. However, we have to consider in this case all rescalings done during the process of the FT calculation --- see below in this section.} of every wave that passed through the measuring system during the considered time period. Notably, Rayleigh surface waves are expected to prevail in the seismic noise in the vicinity of the Virgo building \cite{singha2020}. Therefore, we can neglect vertical differences between the spatial positions of the seismometers ($z$ axis). 

Remarkably, space-time Fourier transform, as any other multi-dimensional Fourier transform, can be calculated in an independent way, cf.\ Eq.\ (\ref{STFT}). That is to say, having the input data of $(x, y, t)$-points, we can calculate, firstly, only the temporal part [obtaining $(x, y, \omega)$-points] and secondly, using this result, we can subsequently calculate the spatial part, obtaining the final result, $(k_x, k_y, \omega)$-points. We may also make that inversely, i.e., calculating first the spatial part and then the temporal part to obtain the final result. 

In the approach presented herein, initially, the average value of the signal was subtracted from the input, for each detector separately. After that, the whole signal was divided into 20-second pieces (time frames), each of them with $N = 10~000$ samples. Thanks to dividing the data into short pieces, we can detect short wave impulses, lasting only a couple of seconds. Due to the fact that the space-time method, as any other Fourier-based method, uses finite sets of data, the questions of spectral leakage and, consequently, the necessity of applying appropriate window functions should be considered, for some artificial components may appear in the final result. For that purpose the Hann window function was used,
\begin{equation}
w(n) = \frac{1}{2}\left[1 - \cos\left(\frac{2\pi n}{N-1}\right)\right],
\label{Hann}
\end{equation}
where $n$ indexes each  measurement in the  $N$-sample time frame. The seismometers are 24-bit and measure voltage changes in the range between $-5$~V and 5~V with sensitivity 77.3 V/m/s. Therefore, the input signal of the seismometers was converted from volts to m/s, by using the appropriate conversion coefficient $5/(2^{23} \cdot 77.3)$.

After the initial preparation of data, explained above, the time-frequency part, for every time frame and for each detector separately, was calculated, using FFT (Fast Fourier Transform) algorithm. After that, the result was multiplied by the appropriate detector response function, which depends on the frequency value $\omega$. Subsequently, the spatial part was calculated, using the values obtained in the first step. To calculate the spatial part in the second step, variables $(x,y)$ were changed to $u=s_f x$ and $v = s_f y$, using the same scaling factor,
\begin{equation}
s_f = \frac{\pi}{\max\left\lbrace |x|_{max},|y|_{max}\right\rbrace},
\end{equation}
so that both spatial dimensions were normalized to $(-\pi, \pi)$ range without distorting the waves. The $(-\pi, \pi)$ range of the spatial data was required by the NUFFT (Non-Uniform FFT) algorithm that was used subsequently. A version of the NUFFT algorithm for two dimensions was applied \cite{dutt1993fast}. NUFFT is the version of FFT algorithm for nonuniform samples, as the locations of the seismometers were not uniform. Finally, after calculating the spatial part, the variables were changed back to $k_x = s_f k_u$, $k_y = s_f k_v$, and three-dimensional complex modulus of the spectrum was calculated to obtain the 3D amplitude spectral density (ASD):
\begin{equation}
\textrm{ASD}(k_x, k_y, \omega) = |\textrm{FT}(k_x, k_y, \omega)|,
\end{equation}
where $\textrm{FT}(k_x, k_y, \omega)$ is given by Eq.\ (\ref{STFT}).

\subsection{Test data}
\label{testdata}

In order to test the algorithm presented above, the following simulation was run. A simulated set of data (adopting the same $(x,y)$ positions of the seismometers) with the superposition of two sinusoidal waves:
\begin{equation}
\psi(x, y, t) = A\cos(k_x x - \omega_1 t) + B\cos(k_y y + \omega_2 t)
\label{superpos}
\end{equation}
is created, where $k_x = 0.578\ \textrm{m}^{-1}$, $k_y = 0.876\ \textrm{m}^{-1}$, $\omega_1 = 3(2\pi)\ \textrm{Hz}$, $\omega_2 = 5(2\pi)\ \textrm{Hz}$, and amplitudes of both waves $A = B = 1000$ V.\footnote{To mimic the seismometer measurements in volts, approximated to integer values, due to the fact that the real seismometer measurements in volts were integer values.}

\begin{figure*}
\begin{center}
\includegraphics[width = \textwidth]{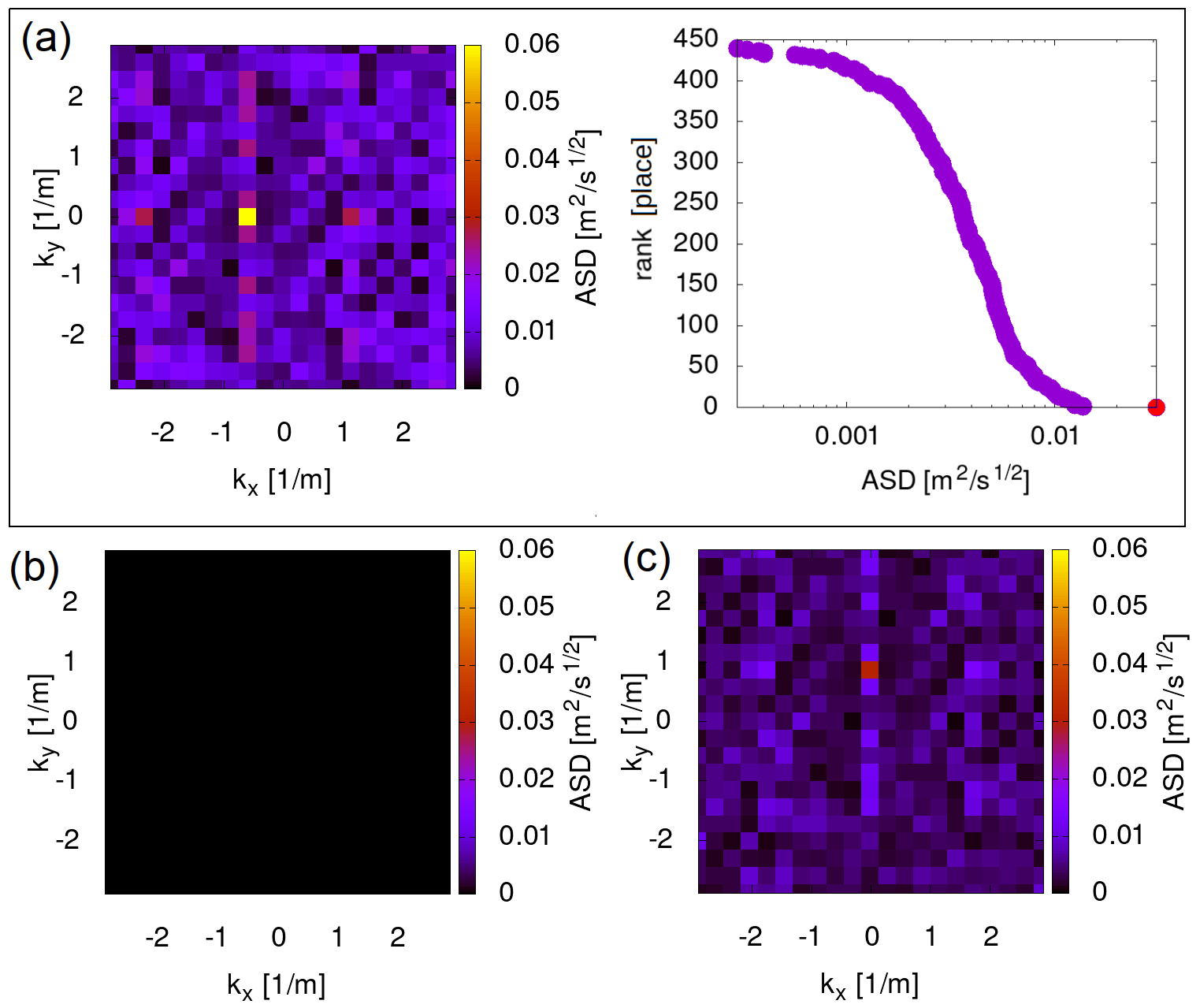}
\caption{Colormap of amplitude spectral density (arbitrary units) given by Eq.\ (\ref{STFT}) for a superposition of two sinusoidal waves given by Eq.\ (\ref{superpos}) (where $k_1 = 0.578\ \textrm{m}^{-1}$, $k_2 = 0.876\ \textrm{m}^{-1}$, $\omega_1 = 3(2\pi)\ \textrm{Hz}$, $\omega_2 = 5(2\pi)\ \textrm{Hz}$, $A = B = 1000$ V) for (a) 3 Hz (on the left; on the right, additionally, a rank of ASD values in semi-log scale is shown, where the outstanding point corresponding to the yellow peak on the left is marked in red), (b) 4 Hz, (c) 5 Hz.}
\label{fig:example}
\end{center}
\end{figure*}

Figure \ref{fig:example} shows the obtained space-time amplitude spectral density for these exemplary data, in a form of colormap, e.g., for 3, 4, and 5 Hz. Additionally, a rank of ASD values is shown for frequency 3 Hz. 
As expected, the space-time amplitude spectral density for such strong signal consists of two outstanding peaks for frequency 3 Hz and 5 Hz [Figs.\ \ref{fig:example}(a) and \ref{fig:example}(c)], corresponding to both parts of the superpositioned wave, with proper wave-vector and frequency values. The peak height for 5 Hz is about two times smaller than for 3 Hz, due to applying detector response function, which partially dampens higher frequencies (cf.\ Sec.\ \ref{method}). However, both peaks are outstanding compared to the other spatial modes for frequencies 3 and 5 Hz, respectively: they are more than 10 times larger than the standard deviation for the whole plot for the particular frequency [on the rank plot for 3 Hz, Fig.\ \ref{fig:example}(a) on the right, the outstanding $\textbf{k}$-value is denoted by the red dot in the bottom right corner]. This is also true for the neighboring frequency values, since both peaks have a wide profile in the frequency domain --- peaks are visible for a large frequency range around 3 and 5 Hz, respectively, with the value of the peak larger than 10 standard deviations of all spatial modes in the plot. However, the result for, e.g., 4 Hz is much weaker than for 3 and 5 Hz, and compared with them shows no signal at all [Fig.\ \ref{fig:example}(b)]. 

Therefore the importance of the ranking-plot representation shown on Fig.\ \ref{fig:example}(a) is that it allows the distinction of the essential points in the amplitude spectral density plot. Consequently, the future signal processing in order to separate the real seismic-wave signal from the noise in the spectral-density data may use the ranking-plot analyses. Nevertheless, for the real data we may expect the obtained peaks of ASD being weaker than in the above test example, due to the fact that the test data contained strong input signal without any additional noise, whereas for the real data the input signal is expected to contain some noise. 

\subsection{Real data}
\label{realdata}

\begin{figure}
\centering
\includegraphics[width=\textwidth]{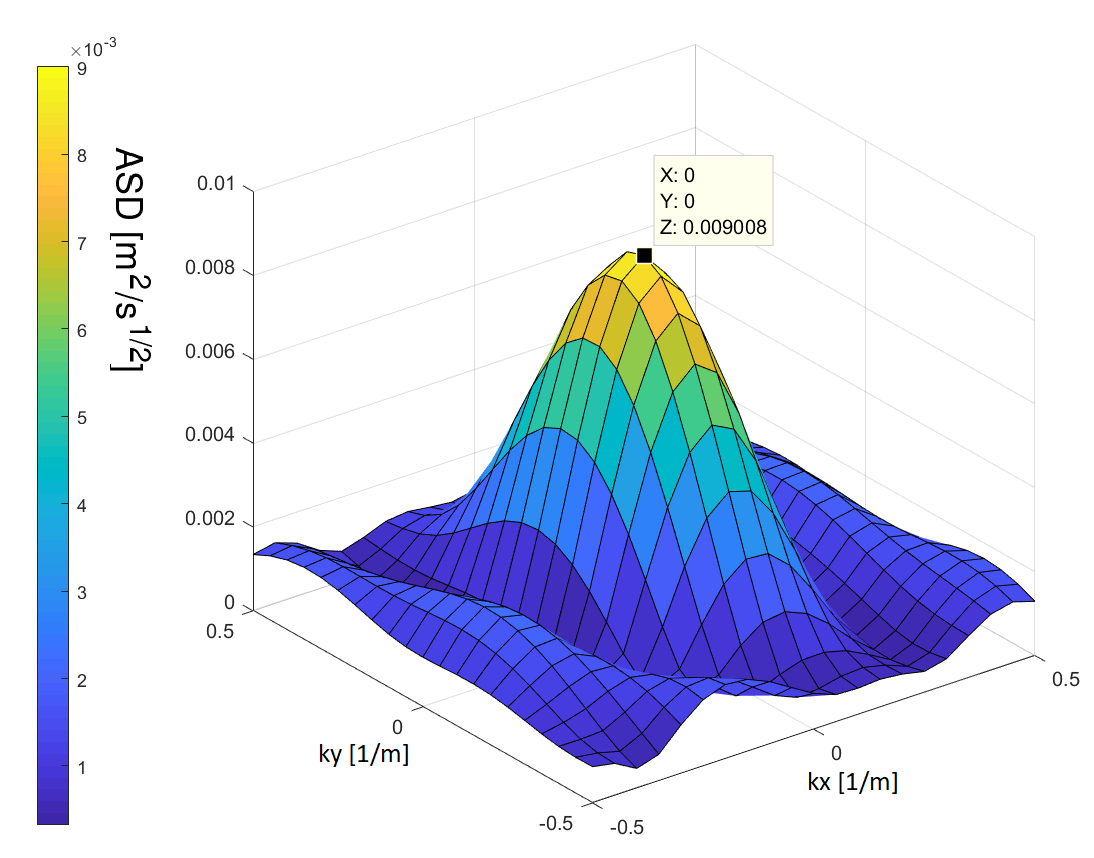}
\caption[cz]{ASD value in $(k_{x}, k_{y})$ space given by Eq.\ (\ref{STFT}) for frequency 5.5 Hz, 10th minute of the first hour of measurements (marked also using colorscale on the left).}
\label{fig:example2}
\end{figure}

After the  successful test of the algorithm on the artificial data, the analysis of the real empirical data collected from the Virgo West-End building was made. In the initial step, the seismometer indications in volts were converted to the appropriate values in m/s. Then, the three-dimensional transforms for these data were calculated using FFT and NUFFT algorithms (cf.\ Sec.\ \ref{method}). Finally, the space-time amplitude spectral density was calculated for each 20-second period and each frequency. An example plot is usually characterized by one main maximum --- cf.\ Fig.\ \ref{fig:example2}.

\begin{figure}
\centering
\includegraphics[width=\textwidth]{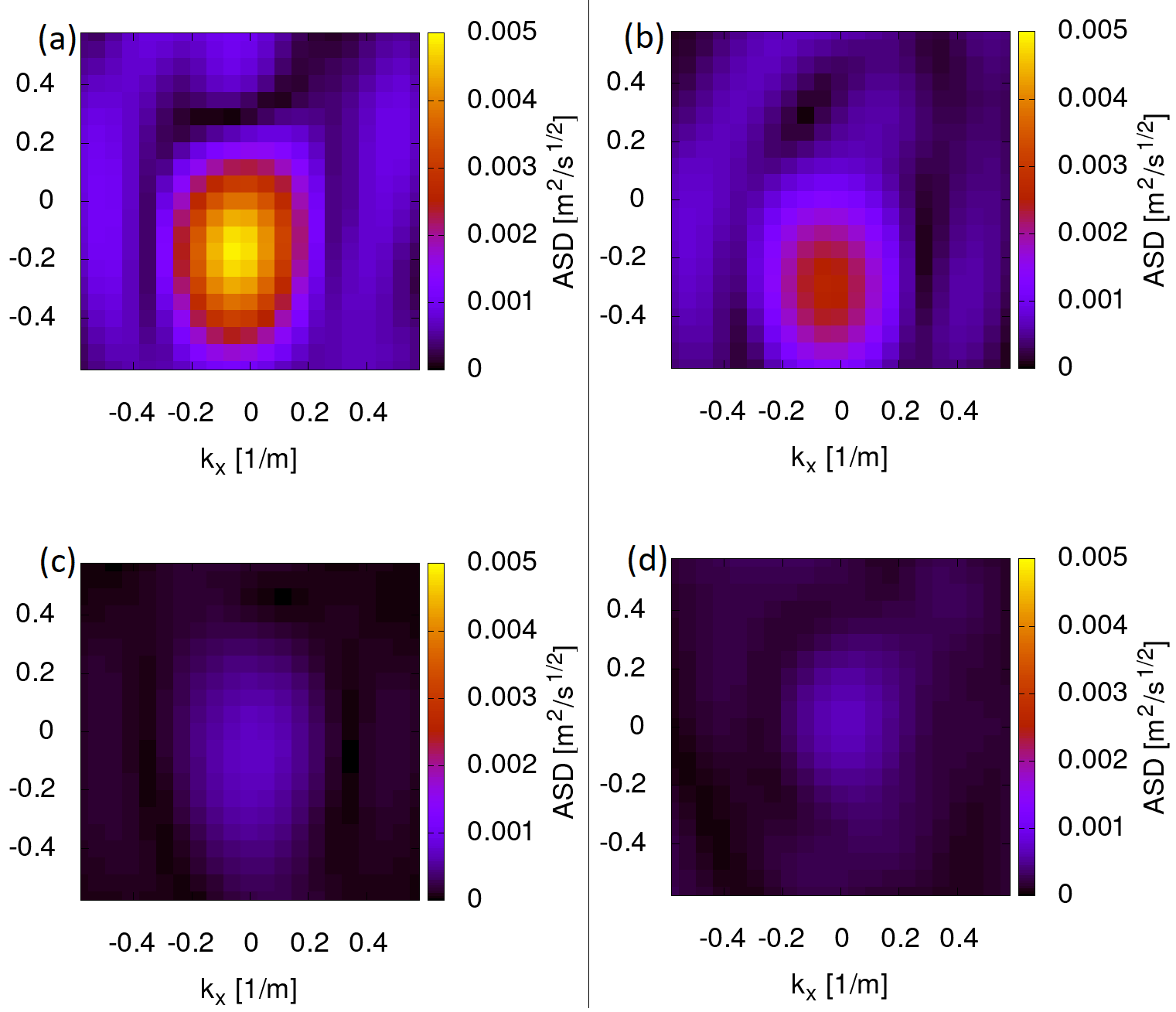}
\caption[cz]{Colormap of ASD in $(k_{x}, k_{y})$ space for 20-second pieces of seismic data from 2018-01-24, 13:30, and for: (a) 5 Hz, (b) 10 Hz, (c) 12 Hz, (d) 15 Hz.}
\label{fig:asd2d}
\end{figure}

The next step was finding a position $(k_{x},k_{y})$ of maximal ASD value in time, for each frequency from the range 5 to 20 Hz, step 0.1 Hz. Examples of preliminary results for 20-second pieces of seismic data, taken from the same hour of measurement, are shown in Fig.\ \ref{fig:asd2d}.
There are dominant waves for particular frequencies, as expected. On first plot for 5 Hz [Fig.\ \ref{fig:asd2d}(a)] there is a strong maximum (north-east wave direction). On Fig.\ \ref{fig:asd2d}(b) for 10 Hz there is much weaker and fuzzy maximum. For (c) and (d) (12 Hz and 15 Hz, respectively) there are not any signals visible, but rather some artifacts present due to inaccuracy of the measurement or the NUFFT method used.

\begin{figure}
\centering
\includegraphics[width=\textwidth]{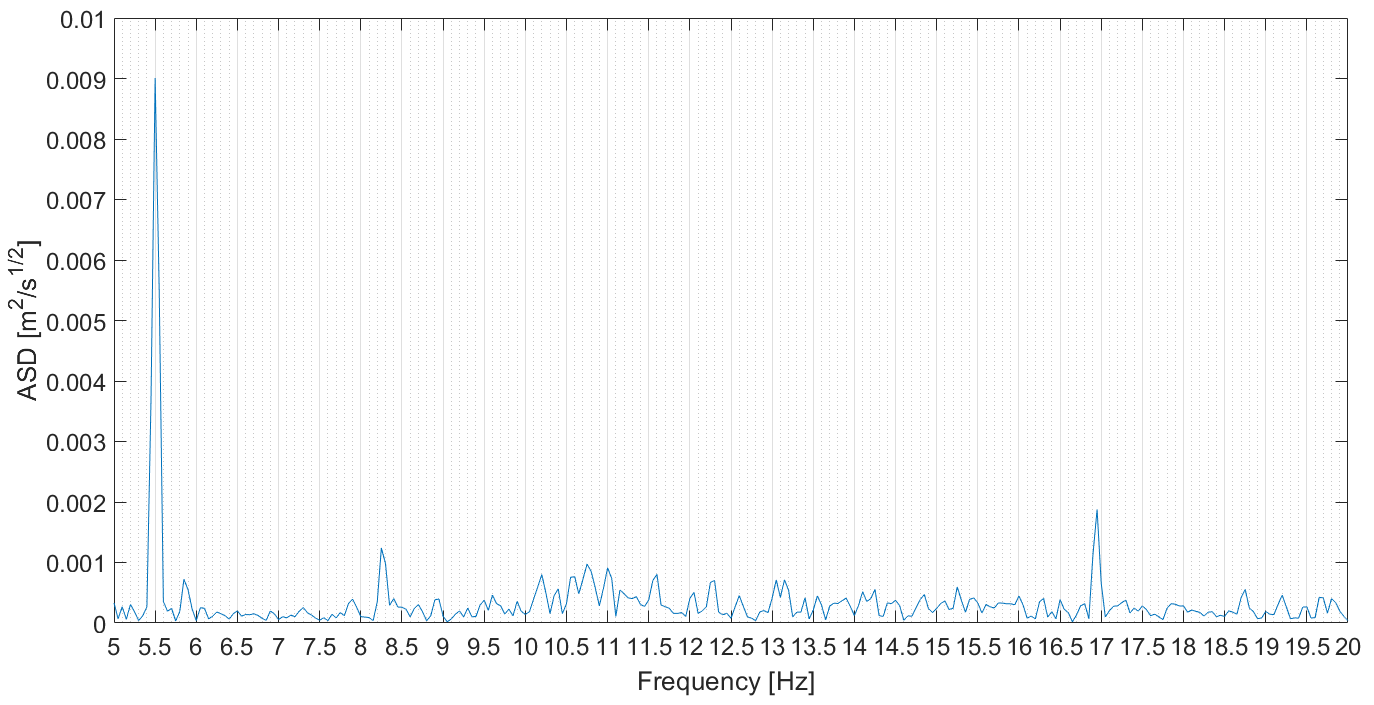}
\caption[cz]{ASD for $(k_x, k_y) = (0,0)$ and frequencies from 5 to 20 Hz, averaged over an example hour of measurement.}
\label{fig:szerf}
\end{figure}

On Fig.\ \ref{fig:szerf} ASD values for $(k_x, k_y) = (0,0)$ and for each frequency, averaged for one example hour of measurement, are shown. Observed maxima of ASD are very sharp, therefore the step 0.1 Hz is taken for this plot. 

\subsection{Dominant waves}
\label{dominant}

Resolution of wave-vector space taken for our calculation is $\delta_k=0.1\ \textrm{m}^{-1}$, thus, for each frequency 180 matrices $21 \times 21$ were obtained. In order to determine dominant waves the following function (paraboloid) was fitted to the space-time Fourier spectra:
\begin{equation}
\textrm{fit}(u,v)=A-\frac{u}{\sigma_{u}}^{2}-\frac{v}{\sigma_{v}}^{2},
\label{fit}
\end{equation}
where A is maximal amplitude for a given matrix $21 \times 21$, $\sigma_{u}$ and $\sigma_{v}$ are standard deviations of $u$ and $v$, respectively, 
\begin{eqnarray}
    u=(k_x-k_{x0})\cdot \cos(\alpha)-(k_y-k_{y0})\cdot \sin(\alpha),
\\   v=(k_x-k_{x0})\cdot \sin(\alpha)+(k_y-k_{y0})\cdot \cos(\alpha),
\end{eqnarray}
$k_{x0}, k_{y0}$ are coordinates of maximal ASD value, $\alpha$ is inclination of the paraboloid axis with respect to the $X$ axis, and $k_x, k_y$ are the matrix coordinates.

In order to increase the resolution in wave-vector space, for each matrix an area of $100 \times 100$ points close to the maximum was inspected, with an accuracy of $\max\{\textrm{ASD}(k_{x},k_{y})\}$ increased by two orders of magnitude. Then, the logarithm, $L$, of the likelihood function:
\begin{equation}
L=\sum_{i,j=1}^{21}\left[\textrm{fit}(k_x,k_y,k_{x0},k_{y0},\sigma_{u},\sigma_{v},\alpha,A)-\lambda(i,j)/\sigma_{L}\right] ^{2}
\label{likelihood}
\end{equation}
was calculated, where $\lambda(i,j)=\textrm{ASD}(k_x,k_y,\omega)$ is a value of ASD-matrix element for a particular $\omega$, and $\sigma_{L}=0.05\cdot A$ is assumed uncertainty of the experiment.\footnote{A change of $\sigma_L$ does not change coordinates of the ASD maximum, just its value.} The values of $\sigma_{u}$ and $\sigma_{v}$ in Eq.\ \ref{fit} were checked in the range 50 to 150 m. After normalization, collected values of $L$ were aggregated in the following way\footnote{RHS of Eq.\ (\ref{sum1}) does not include $k_x, k_y$ as they reflect indices of summation $i, j$ in Eq.\ (\ref{likelihood}), and $A$, as it was constant for all calculated $L$ values.}
\begin{eqnarray}
L_{x,y}(k_{x0},k_{y0})=\sum\limits_{\sigma_{u},\sigma_{v},\alpha}L(k_{x0},k_{y0},\sigma_{u},\sigma_{v},\alpha)
\label{sum1}
\\L_{x}(k_{x0})=\sum\limits_{k_{x0}}L(k_{x0},k_{y0})
\\L_{y}(k_{y0})=\sum\limits_{k_{y0}}L(k_{x0},k_{y0})
\end{eqnarray}
Finally, the cumulative distribution function and the coordinates of ASD maximum for each frequency and each 20-second period of the measurement were obtained.

\section{Results}
\label{results}

\begin{figure}
\centering
\includegraphics[width=\textwidth]{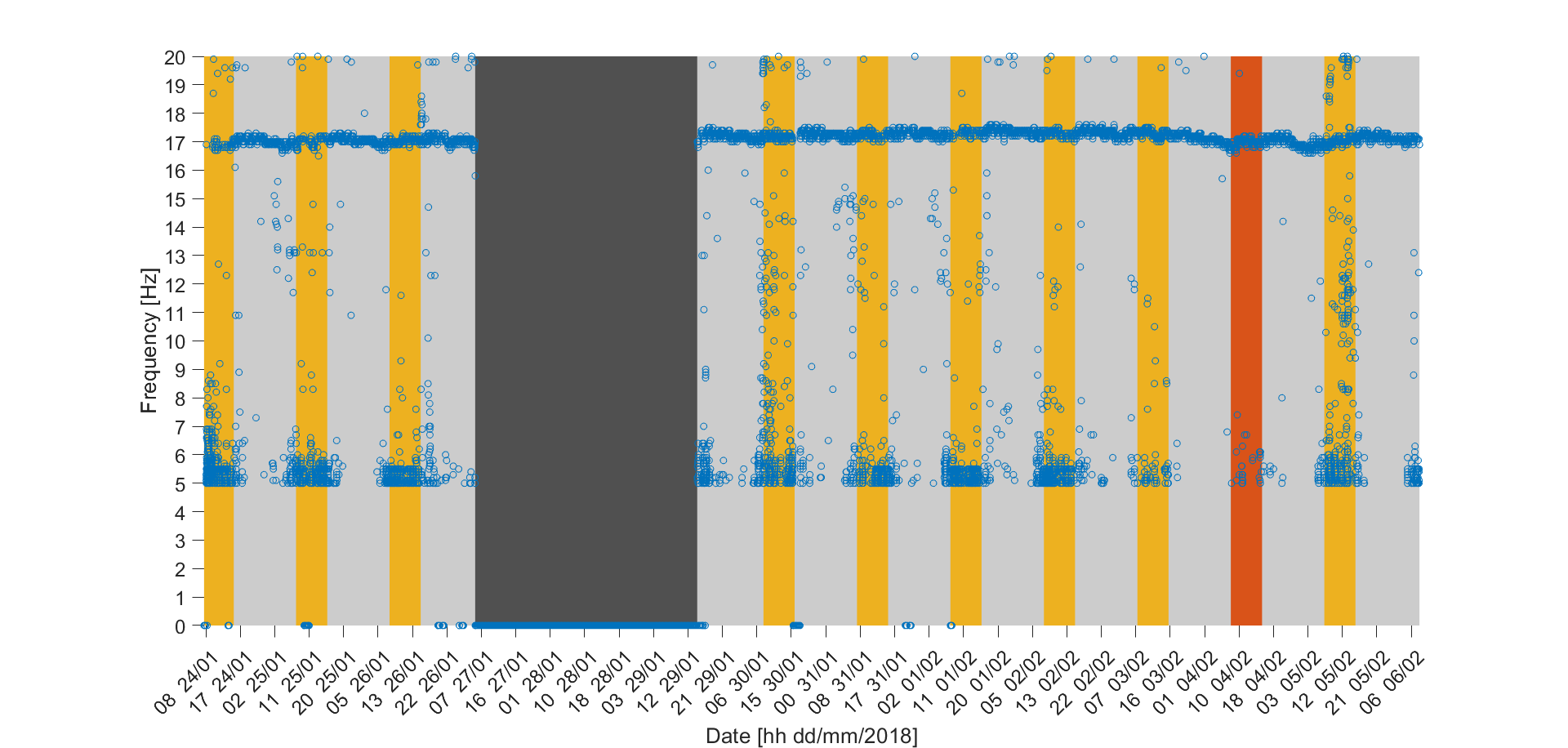}
\caption{Dominant frequency for each time moment. The hours between 8 a.m. and 4 p.m. each day are indicated with yellow color, sundays are indicated with orange, dark grey is a period of measurements break.}
\label{fig:freq}
\end{figure}

\begin{figure}
\centering
\includegraphics[width=\textwidth]{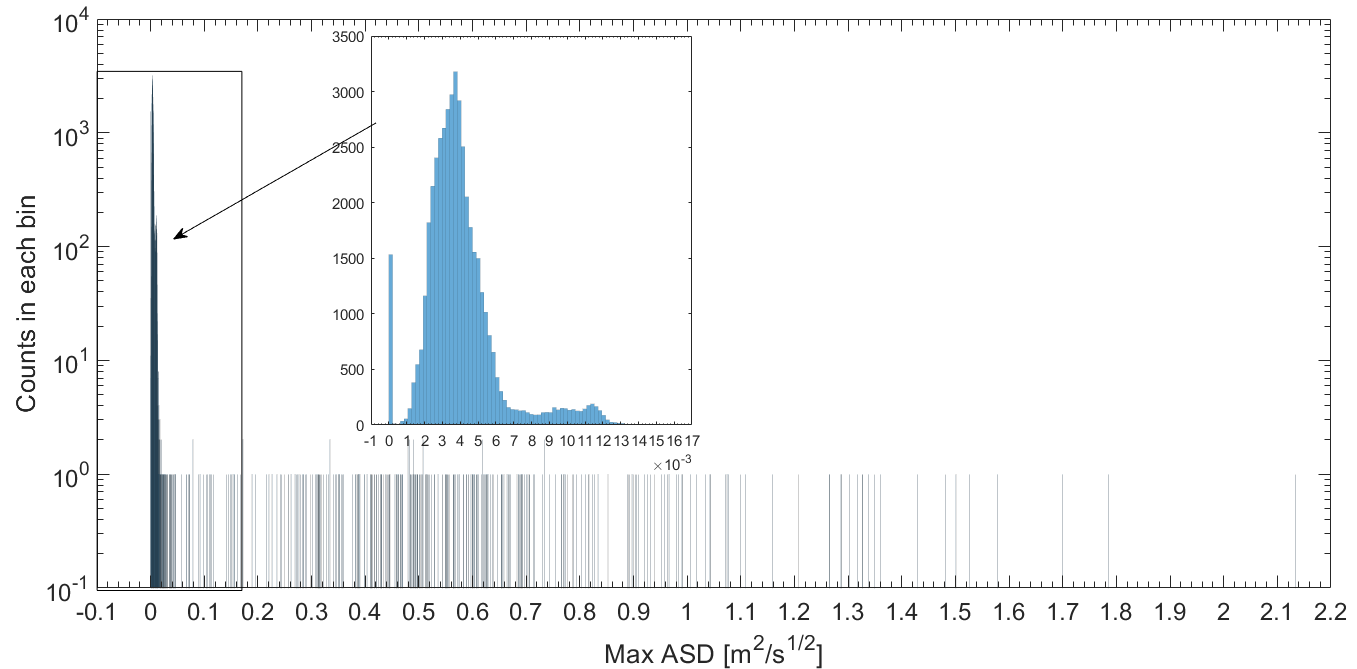}
\caption{Histogram of maximal ASD values (bin width $2\cdot10^{-4}$).}
\label{fig:asd}
\end{figure}

\begin{figure}
\centering
\includegraphics[width=\textwidth]{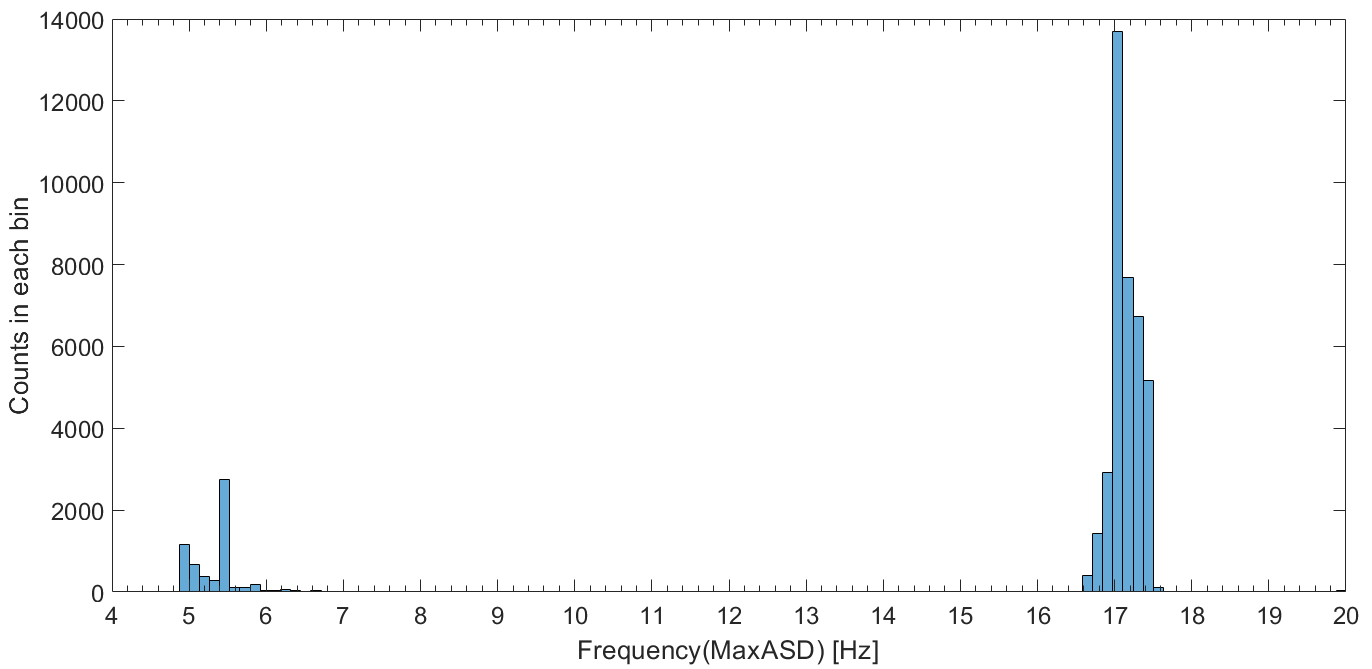}
\caption{Histogram of dominant frequencies (bin width 0.13).}
\label{fig:histf}
\end{figure}

In this section the final results for all data gathered between 24-01-2018 and 06-02-2018 are presented. Fig.\ \ref{fig:freq} shows dominant frequencies for the whole period. 
Apparently, there are two main frequencies for ASD: ca.\ 5-6 Hz and ca.\ 17 Hz. The first one is characterized by strong day-night periodicity. On Figs.\ \ref{fig:asd} and \ref{fig:histf} histograms of the data are shown. Fig.\ \ref{fig:asd} shows a distribution of max(ASD) values. Fig.\ \ref{fig:histf} shows a distribution of dominant frequencies.\footnote{Counts for zero value on both plots are a consequence of quality of the data and should not be considered at all --- cf.\ Fig.\ \ref{fig:quality}.} Two main dominant frequencies are 5.5 Hz and 17.1 Hz. 

\begin{figure}
\centering
\includegraphics[height=\textwidth, angle=270]{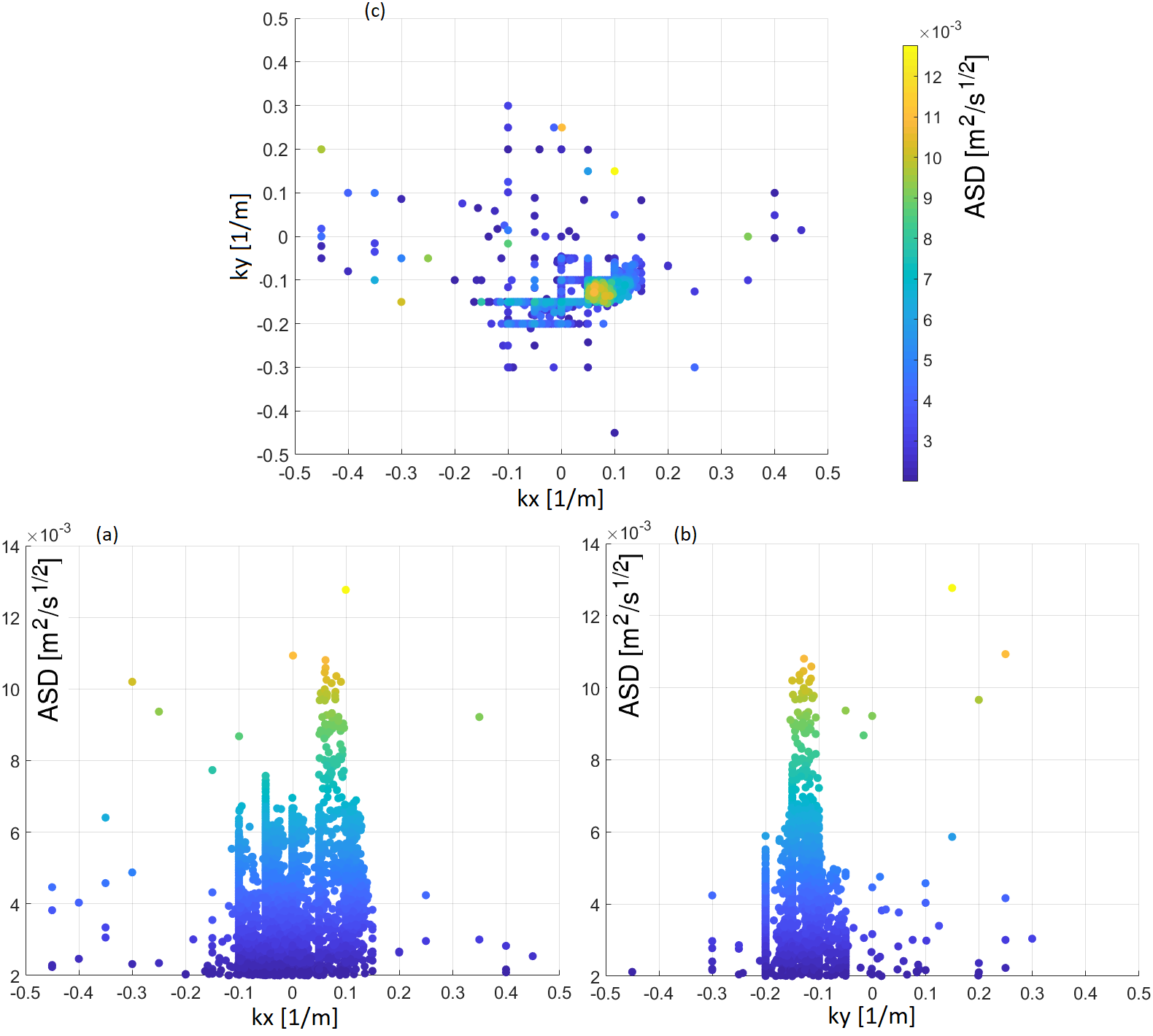}
\caption{Locations of max(ASD) in wave-vector space for frequency 17.1~Hz projected onto (a) $XZ$, (b) $YZ$, (c) $XY$ surface where $X$ and $Y$ reflect $k_x$ and $k_y$, respectively, and $Z$ reflects ASD. Colors indicate ASD values.}
\label{fig:171}
\end{figure}

\begin{figure}
\centering
\includegraphics[height=\textwidth, angle=270]{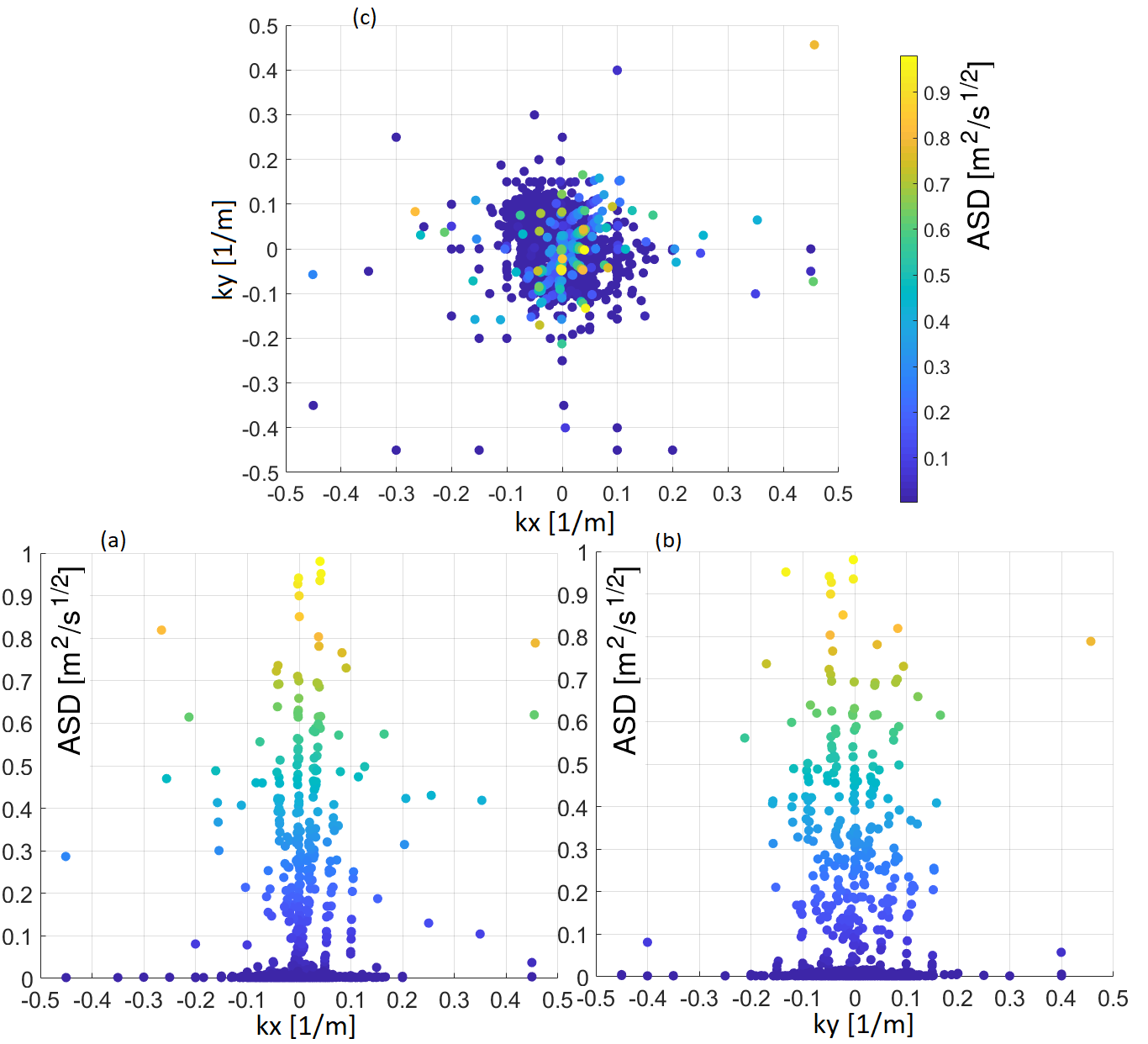}
\caption{Locations of max(ASD) in wave-vector space for frequency 5.5~Hz projected onto (a) $XZ$, (b) $YZ$, (c) $XY$ surface where $X$ and $Y$ reflect $k_x$ and $k_y$, respectively, and $Z$ reflects ASD. Colors indicate ASD values.}
\label{fig:55}
\end{figure}

In the next step, a detailed analysis was made in order to characterize seismic waves in the vicinity of the Virgo WEB. Using methodology from Sec.\ \ref{dominant} exact coordinates of max(ASD) in the wave-vector space, for particular frequencies and 20-second periods of the measurement, were found. Collected results are shown on Fig.\ \ref{fig:171} for 17.1 Hz and Fig.\ \ref{fig:55} for 5.5 Hz. 
For the sake of clarity, the results are projected on $XZ$, $YZ$, and $XY$ surfaces, where $X$ and $Y$ reflect $k_x$ and $k_y$, respectively, and $Z$ reflects ASD. Due to presence of noises in the measurement system and inaccuracy of our method, a lower limit for max(ASD) was set to $0.5\%$ of the maximal ASD value for all data, which was equal to $2.13~\textrm{m}^2/\textrm{s}^{1/2}$. 

\begin{figure}
\centering
\includegraphics[width=\textwidth]{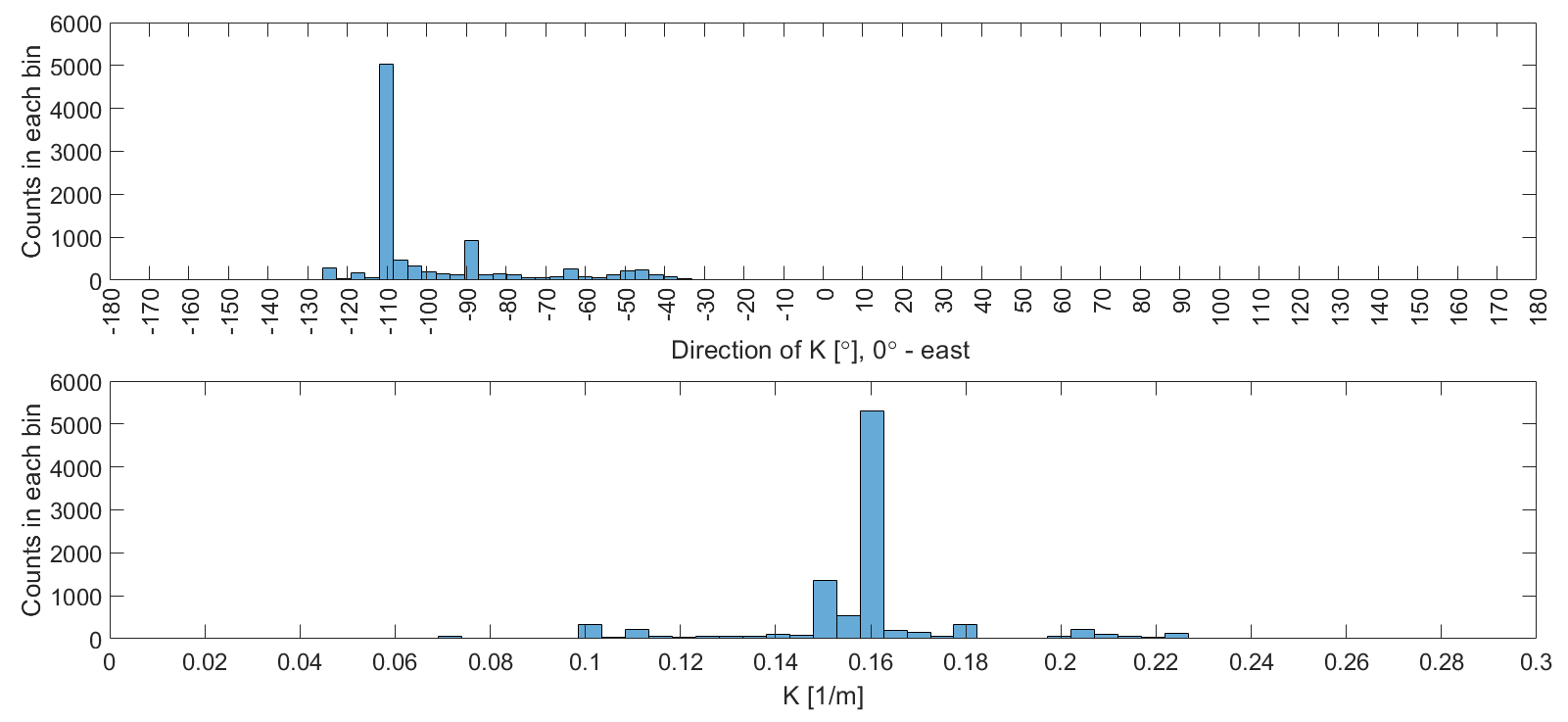}
\caption{Histograms of direction (top) and module (bottom) of the wave vector for frequency 17.1 Hz. Bin width: $3.6^o$ (top), $5\cdot10^{-3}$ 1/m (bottom).}
\label{fig:171kier}
\end{figure}

\begin{figure}
\centering
\includegraphics[width=\textwidth]{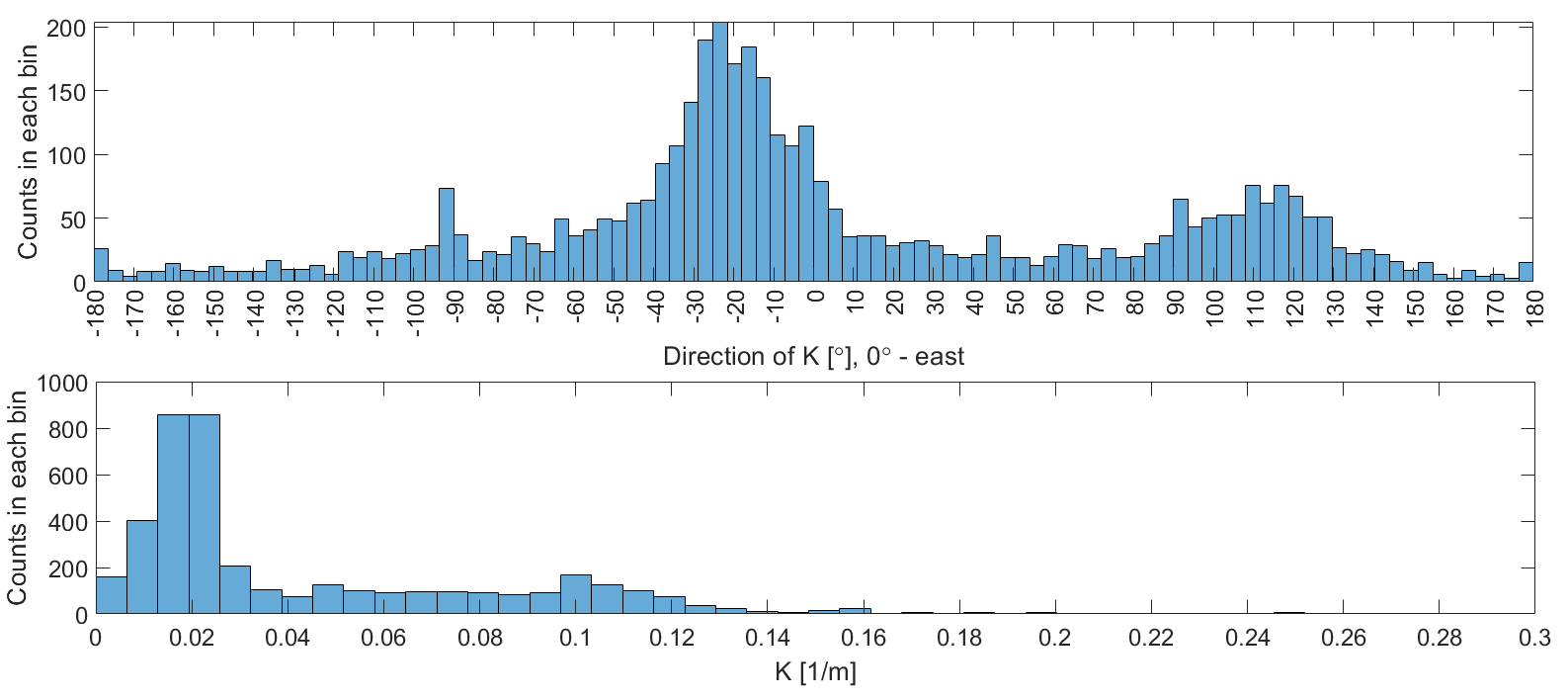}
\caption{Histograms of direction (top) and module (bottom) of the wave vector for frequency 5.5 Hz. Bin width: $3.6^o$ (top), $5\cdot10^{-3}$ 1/m (bottom).}
\label{fig:55kier}
\end{figure}

Figs.\ \ref{fig:171kier} and \ref{fig:55kier} for frequencies 17.1 Hz and 5.5 Hz, respectively, are histograms of wave direction (upper parts) and wavenumber (lower parts). $0^o$ on upper plots does not indicate cardinal direction, but direction of $X$ axis in the seismometer coordination system shown on Fig.\ \ref{fig:latt}.
For frequency 17.1 Hz results indicate clearly the direction $-110\pm2.5^o$ and wavenumber $|K_{17.1}|= 0.158$. These values correspond to the seismic wave arriving from northeast with velocity $680 \pm 111$ m/s. Results for 5.5 Hz are more ambiguous and finding a strict wave direction in this case is impossible. The most frequent direction is ca.\ $-25\pm16.3^o$ with wavenumber $|K_{17.1}|= 0.02$. These values correspond to the seismic wave arriving from northwest with velocity $1727 \pm 431$ m/s. 

\begin{figure}
\centering
\includegraphics[width=\textwidth]{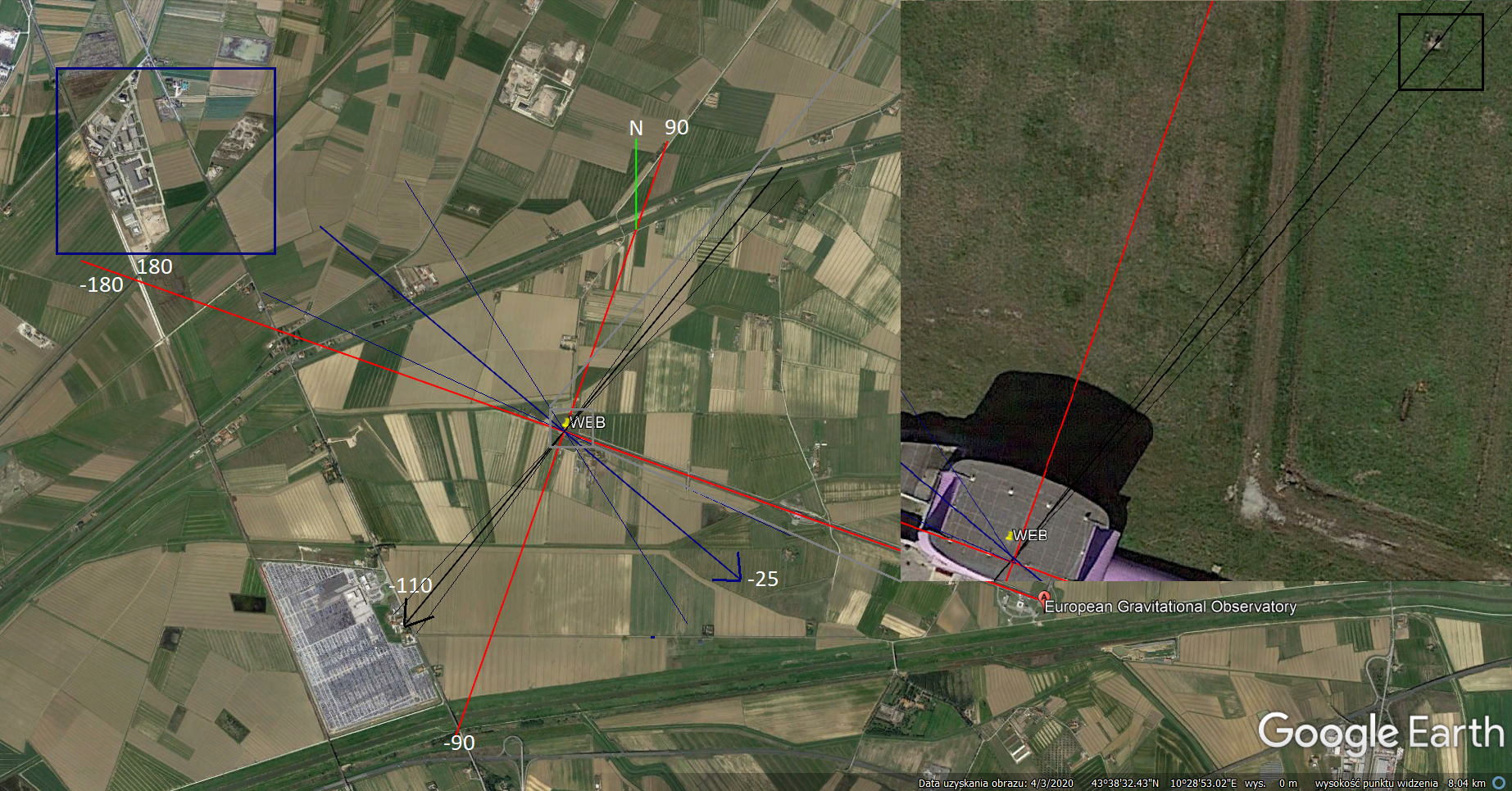}
\caption{Map of the vicinity of the Virgo WEB ($8 \times 4$ km). Red lines indicate coordination system consistent with that shown on Fig.\ \ref{fig:latt}. The directions of waves propagation for 5.5-Hz wave with error $\pm 16.3^o$ (blue arrow + side lines) and for 17.1-Hz wave with error $\pm 2.5^o$ (black arrow + side lines). Rectangles indicate the possible wave sources. The area close to the 17.1-Hz wave source is zoomed and shown in the top right corner of the figure.}
\label{fig:mapa}
\end{figure}

On Fig.\ \ref{fig:mapa} the map of the vicinity of the Virgo detector with directions of waves propagation is shown. Red lines is an assumed coordination system, shifted $19.5^o$ clockwise to true north. Blue lines indicate direction of 5.5-Hz wave with errors. The area in blue rectangle, i.e., an industrial complex 2.5 km away from the WEB, may be a source of this wave. Black color indicate direction of 17.1-Hz wave. In top right corner of Fig.\ \ref{fig:mapa} a zoom of the vicinity of the WEB is shown, with a potential source of 17.1-Hz wave marked (black rectangle), ca.\ 100 m away for the WEB.

\section{Summary and discussion}
\label{summary}

\subsection{Conclusions}

Using the data from the experiment conducted between 24-01-2018 and 06-02-2018 collected by the system of seismometers inside the west-end building of the Virgo gravitational interferometer, we made a seismic analysis of the nearby area. The seismic noise (currently compensated) corresponds to the changes in the gravitational field that can be detected by the Virgo. These fluctuations should be also identified in order to increase sensitivity of the GW measurement system. Particularly, for a frequency below 20-30 Hz they are recognized as newtonian noise \cite{harms2015newtonian, Harms, 20Hz}. 

In this work, we searched for the seismic waves with a frequency from 5 to 20 Hz. Two dominant frequencies were detected and identified with two main sources, the first one emitting a continuous wave, and the second one emitting wave cyclically. First observed dominant frequency is 17.1 Hz. The direction of this wave is $-110\pm2.5^o$, i.e., it was arriving from northeast, with velocity $680 \pm 111$ m/s. On Fig.\ \ref{fig:mapa} in the top right corner a small object, ca.\ 100 m away from the WEB is shown. It is located exactly at the found wave direction and can be a source of 17.1-Hz wave. 

However, the source of the 17.1-Hz wave can be also internal, located inside the building. Therefore, another possible interpretation of this wave is a resonance of the whole building, which is excited by vacuum pumps, cooling fans, or other devices  \cite{tringali2019seismic}.

Seismic waves at the frequency 5.5 Hz coincide with daily human activity. They dominate between 6 a.m. and 6 p.m. every day and move with velocity $1727 \pm 431$ m/s. A direction of this waves is not strict, ca.\ $-25\pm16.3^o$. This is caused by a supposed origin of the waves, which is human activity in the nearby industrial complex. The study made before, indicated another antropogenic source of the noise at the frequency 4-10 Hz: three nearby bridges \cite{mosty}. The velocities given above are typical for dispersion media, i.e., the wave with a lower frequency moves faster, and agree with the predictions of one of the models from \cite{hughes1998seismic}. The typical velocity of Rayleigh waves in metals is 2--5 km/s, while in soil 50--700 m/s for shallow waves (under 100 m) and 1.5--4 km/s for depth over 1 km \cite{vel}. 

\subsection{Future work}

The method of seismic-wave detection in the Virgo WEB presented above, is able to detect particular waves that pass through the building. The main advantages of the method is that it gives the full information about the waves that come through the measuring system (frequency, length, direction, and amplitude) with low loss of the information from the seismometers comparing, e.g., to the method using coherence values between each pair of the detectors during a certain measurement period, for example, one hour \cite{harichandran1986stochastic, harms_coherence, tringali2019seismic}. Our method is more sensitive and efficient, and it is most useful for the characterization of seismic field, as it was shown in this work.

The idea of minimization of newtonian noise in the Virgo GW signal is based on monitoring sources of gravitational field perturbations. These perturbations are caused by local changes of density in the vicinity of the mirrors. A common method of filtering newtonian noise is based on Wiener filters, which have been already applied to reduction of other noises in GW detectors. The parameters of seismic waves found in this work can be used to fit correlation models of the Wiener filter \cite{nncancel}. 

As the system of mirrors in the Virgo is complex (see Fig.\ \ref{fig:Virgo_schem}), the analysis conducted herein shall also be made for other parts of the detector. For the west-end building itself the data were collected for more than a year; in the north-end building a system of sensors is going to be installed as well. Complementarily to the system of seismometers, infrasound microphones are going to be used for a precise determination of the sources of gravitational field fluctuations caused by pressure changes. Particularly, a detailed analysis of the frequencies below 10 Hz would be vital, as for 1.7 Hz a significant noise coming from the nearby wind farm is observed \cite{saccorotti2011seismic}. 

One of the problems that needs solving when we use the space-time spectral analysis method is how to distinguish between the real waves and the noise. For instance, some threshold values may be used for this purpose \cite{denys2016universality, schafer1979space}. However, in this article we focused on the dominant frequencies observed in the obtained signal. Moreover, some problems with the spatial aliasing in the results should be resolved, e.g., by applying the window function also for the spatial dimensions (cf.\ Sec.\ \ref{method}). On top of that, we may consider the use of overlapping time periods when calculating the spectra, to improve the accuracy and relevance of the results. The further studies will contribute to the improvement of quality of the information about the gravitational-wave sources like binary black holes and neutron stars.

\section*{Acknowledgments}

M.D. is grateful for inspiring discussion with Krzysztof Lorek and Jan Harms. The authors are grateful to colleagues from the Virgo collaboration and the Virgo Newtonian noise team for providing the seismic array data and for fruitful discussions. The work was supported by the TEAM grant from the Foundation for Polish Science TEAM/2016-3/19 and by the grant \emph{AstroCeNT --- Particle Astrophysics Science and Technology Centre International Research Agenda} carried out within the International Research Agendas programme of the Foundation for Polish Science financed by the European Union under the European Regional Development Fund.

\begin{appendices}

\section{Coordinates of the seismic sensors}
\label{appendix}

\begin{table}[h]
\small
\begin{center}
\begin{tabular}{|c|c|c|c|c|c|c|c|}
\hline
No. & $x[m]$ & $y[m]$ & $z[m]$ & No. & $x[m]$ & $y[m]$ & $z[m]$ \\
\hline
        1 & -2996.1855 & 3.3109 & -3.4011 & 20 & -3003.6825 & -6.6868 & -3.4195 \\ \hline
	    2 & -2998.6876 & 7.1706 & -3.4071 & 21 & -3000.1694 & -3.4201 & -3.4149 \\ \hline
	    3 & -2999.3596 & 3.2421 & -3.4008 & 22 & -2999.4935 & -5.4172 & -3.4146 \\ \hline
	    4 & -3003.9149 & 7.8757 & -3.4227 & 23 & -2996.6250 & -3.3377 & -3.4085 \\ \hline
	    5 & -3003.0811 & 3.2749 & -3.4150 & 24 & -2999.3105 & 2.7366 & -3.3156 \\ \hline
	    6 & -3005.9728 & 7.7190 & -3.4288 & 25 & -3002.9757 & 2.6436 & -3.3247 \\ \hline
	    7 & -3006.2434 & 3.3279 & -3.4237 & 26 & -3006.8761 & 2.7944 & -3.3304 \\ \hline
	    8 & -3010.3192 & 7.9888 & -3.4437 & 27 & -3010.9353 & 2.8204 & -3.3416 \\ \hline
	    9 & -3012.1375 & 3.3488 & -3.4299 & 28 & -3013.8210 & 2.3477 & -3.3491 \\ \hline
	    10 & -3014.7164 & 7.5541 & -3.4490 & 29 & -3013.6882 & -2.6371 & -3.3435 \\ \hline
	    11 & -3017.8701 & 4.2944 & -3.4613 & 30 & -3011.2803 & -2.5983 & -3.3446 \\ \hline
	    12 & -3014.9770 & -0.0874 & -3.4426 & 31 & -3006.4946 & -2.6779 & -3.3386 \\ \hline
	    13 & -3017.9104 & -4.0994 & -3.4595 & 32 & -3002.6733 & -2.5689 & -3.3296 \\ \hline
	    14 & -3015.1017 & -7.3702 & -3.4573 & 33 & -2999.6753 & -2.5736 & -3.3223 \\ \hline
	    15 & -3013.7171 & -3.1819 & -3.4287 & 34 & -2999.8896 & 0.0321 & -3.3182 \\ \hline
	    16 & -3011.1778 & -6.0311 & -3.4400 & 35 & -3009.4183 & 0.0789 & -3.3497 \\ \hline
	    17 & -3009.9611 & -3.1498 & -3.4302 & 36 & -3013.7110 & -0.0613 & -3.3462 \\ \hline
	    18 & -3005.5196 & -6.1313 & -3.4285 & 37 & -3012.2423 & -1.1159 & -6.8445 \\ \hline
	    19 & -3005.9568 & -3.1540 & -3.4234 & 38 & -3008.4083 & 1.4434 & -6.8571 \\ \hline
\end{tabular}
\caption{Coordinates of the seismometers with respect to the Virgo test mass.}
\label{tab_coords}
\end{center}
\end{table}

\end{appendices}

\FloatBarrier

\bibliographystyle{unsrt}

\end{document}